\documentclass[aps,prb,amsmath,a4paper,graphicx,amsfonts]{revtex4}
\usepackage{graphics}
\usepackage{amsfonts}
\usepackage{graphicx}
\usepackage{epic,eepic,epsfig}
\usepackage{dcolumn}
\usepackage{bm}
\begin{document}
\begin{center}
\vspace{5cm} {\Large {\bf Toda lattice representation for  random
matrix model with logarithmic confinement}}
\vspace{5cm}

{\bf T.~A.~Sedrakyan
}\\

{\it
The Abdus Salam International Centre for Theoretical Physics,
Strada Costiera 11, 34100 Trieste, Italy \\
email: tsedraky@ictp.it}\\
\date{\today}
\vspace{3cm}
\end{center}
%\begin{abstract}
\begin{center}
{\bf Abstract}
\end{center}
We construct a replica field theory for a random matrix model with
logarithmic confinement  [K.~A.~Muttalib  et.al., Phys. Rev. Lett.
71, 471 (1993)]. The corresponding replica partition function is
calculated exactly for any size of matrix $N$.  We make a
color-flavor transformation of the original model and find
corresponding Toda lattice equations for the replica partition
function in both formulations. The replica partition function in
the flavor space is defined by generalized Itzikson-Zuber (IZ)
integral over homogeneous factor space of pseudo-unitary
supergroups $SU(n\mid M,M)/SU(n\mid M-N,M)$ (Stiefel manifold)
with $M\rightarrow\infty$, which is evaluated and represented in a
compact form.
%\end{abstract}

%\pacs{72.15.Rn, 02.50.-r, 05.40.+j}
%\keywords{}
%\maketitle

\newpage
\section{Introduction}

\noindent Recently a considerable interest appeared towards
analyzing of integrable structure behind Gaussian random matrix
theories (RMT)~\cite{Meh}. An important development in this
context was made by Kanzieper in \cite{Kanz1}, where the link
between nonlinear replica $\sigma$-models and Painlev$\acute{e}$
hierarchy of exactly solvable Toda lattice equations~\cite{Toda}
was established. More precisely it has been shown, that the
replica partition function of the Gaussian  unitary ensemble
(GUE), which in it's original~\cite{EA} and dual (in the replica
space) $\sigma$-model~\cite{Weg},\cite{ELK} representation
satisfies  the Toda lattice equation (TLE),  can be reduced to
Painlev$\acute{e}$ transcendent $\varphi_{IV}$~\cite{FW}. From
here exact nonperturbative results were obtained demonstrating the
ability of the method to overcome the critique of replica trick
\cite{Zirn2}. In the chiral unitary ensemble (chUE), the replica
trick on the basis of TLE was analyzed in~\cite{Kanz1,SV1}.

In this paper we show how the above picture may arise and work in
a unitary ensemble defined by probability distribution function
$P(H)=\exp[-tr V(H)]$ with the confining potential
\begin{eqnarray}
\label{Pot}
V(x)=\sum_{n=0}^{\infty}\ln [1+2q^{n+1}\cosh \beta\chi +q^{2n+2}],
\end{eqnarray}
that behaves as  $V(x)\propto \ln^{2}(|x|)$ at $|x|\rightarrow\infty$.
Here $x=\sinh \beta\chi/2$ and the parameter $q=e^{-\beta}$, with
$\beta > 0$.

Defined and solved using the orthogonal polynomials
method in~\cite{Muttalib}, this unitary random matrix model has
attracted considerable attention during the last decade.
One reason for that is that it leads to the eigenvalue distribution
which for not very large $\beta < \pi^{2}$ is intermediate between the
Wigner-Dyson ~\cite{Meh} and the Poisson distribution and has the
features typical of the critical level statistics ~\cite{KM,KT}.

Another reason, perhaps more important, is  that the
potential  Eq.(\ref{Pot}) fits very well the function 
$a [\ln(1+b x^2)]^2$ in a wide region of the variable $x$.
%has a log-normal behavior 
%for a wide region of the variable $x$.
%fitted very well with the function $a [\ln(1+b x^2)]^2$
%for a wide region of the variable $x$. 
This behavior was obtained
numerically in the %qasi-one dimensional 
tight binding Anderson model with random site energies 
as a distribution potential of variables 
%for the Transfer martix eigenvalue 
$x_i$, $i=1,\dots,N$, 
%as a potential for the distribution of the variables
%$x_i$, 
connected with the conductance as $g= \sum_{i=1}^N (1+x_i^2)^{-1}$
~\cite{Muttalib2}, reflecting a log-normal behavior
of the conductance $g$ in a product of random matrices ~\cite{Imry}.
Therefore, by treating
matrices $H$ as $H^2= (T T^+ + T^+ T -2I)/4$, where $T$ is the transfer
matrix characterizing the disordered conductor,  the model with the
potential (\ref{Pot})
can be considered as a qualitatively correct phenomenological model
for the Anderson insulator
with the parameter $\beta$ playing a role of the ratio of the
system size to the localization length $L/\xi$ and $\chi/\xi$ playing a
role of the Lyapunov exponent ($I$ is an identity matrix). At large $\beta$
the probability density $\rho(\chi)$ is strongly peaked ~\cite{Muttalib},
\cite{Bogomol} near positive
integers exhibiting the phenomenon which is  known in
theory of quasi-1d Anderson
localization as transmission eigenvalue {\it crystallization}
\cite{Muttalib3}, \cite{PZIS}, \cite{BR},
\cite{Lama}.

Up to now the model Eq.(\ref{Pot}) was investigated only within the
approach of orthogonal polynomials which was successful because the
corresponding set of orthogonal polynomials (q-deformed Hermite
polynomials)
is explicitly known \cite{Ismail}. Though this method gives an exact
solution to the problem it does not tell us anything about the connection
of this problem to exactly integrable lattice models. On the other hand,
such a connection is expected by
analogy with the Wigner-Dyson RMT which is solvable by the ordinary
Hermite polynomials. Thus the main question we ask is as to whether or
not
the RMT
Eq.(\ref{Pot}) is related with exactly integrable lattice models and what
precisely these models are.

As a first step we compute
the replica partition function exactly using the pole structure of
the distribution function $P(H)$. Indeed, the form of the potential
allows us to integrate over poles in complex plane and obtain the
new exact expression for the partition function as an infinite sum of
residues.
Such a result of nonperturbative nature turns out to be
exactly deducible from the dual
representation of the partition function in a form of
the Itzikson-Zuber (IZ) type matrix integral over homogeneous
factor space of pseudounitary supergroups $\frac{SU(n\mid M,M)}
{SU(n\mid M-N,M)}$ (Stiefel manifold) with $M \rightarrow\infty$.
Thus one of the results of the paper is a calculation of this specific
IZ integral by use of Duistermaat-Heckman theorem.
To the best of our knowledge this integral was not
calculated earlier in the literature.

One of the main results of the present paper is the establishment of the
link of RMT Eq.(\ref{Pot}) with integrable lattice models. For the
original replica partition functions we infer the same Toda
lattice equations as in the case of the Gaussian RMT \cite{Kanz1}.
However the dual partition functions $Z_{n,N}$ obtained by a
super-symmetric mapping onto a variant of the replica $\sigma$
model (color-flavor transformation~\cite{Zirn}) has a more
delicate relationship with the Toda lattice models. Namely,
semi-infinite hierarchies of fermionic and bosonic flavors
involved in this mapping manifest themselves as an extension
of graded TLE first found in \cite{SV2} in the context of chiral
unitary ensemble relevant for QCD.

Finally, we analyze the new representation for the two-level correlation
function $R_{2}(x)$ and the probability density $\rho(\chi)$ and discuss
the phenomenon of $\chi$--crystallization.

The paper is organized as follows. In the second section we
make a color-flavor transformation in the model under consideration
and give its sigma-model formulation defined on the homogeneous factor
space of pseudo-unitary supergroups.
We calculate the replica
partition function in both, color and flavor spaces respectively.
In the third section
 we infer Toda lattice equations in both spaces. In the fourth section,
 by use of
the method of pole integration, we calculate exactly the density of
states and the
two point correlation function at large N. Comparison with known
results follows. Section 5 contains our conclusions.
The calculation of the Itzikson-Zuber matrix integral over Stiefel
manifold and
the derivation of TLE in flavor space are presented in two Appendices.
\section{The replica partition function.}

\noindent
%{\it The replica partition function.}---
The replica partition function of the model Eq.(\ref{Pot}),
defined by $Z_{n,N}(\epsilon)=\langle \det ^{n}(\epsilon
-H)\rangle _{H}$, can be represented as an integral
\begin{eqnarray}
\label{Z1} (C_M)^N\int d H {\det}^{n}(\epsilon -H)\prod _{k=1}^{M}
\frac{1}{\det \left[(\mu _k + i H)(\mu _k - i H)\right]},
\end{eqnarray}
where $\mu _k=\cosh {k\beta \over 2}$,
$C_M=(\prod_{n=1}^M4q^{n})^{-1}$ and the limit
$M\rightarrow\infty$ is understood.
The main idea of the replica trick
is the possibility of analytic continuation of $Z_{n,N}$ to
non-integer values of $n$, which was shown to be exact in GUE
\cite{Kanz1}.

Diagonalizing Hermitian $N\times N$ matrices $H=V\;
diag\{E_{i}\}\; V^{\dagger}$ and taking into account the
Vandermonde determinant one can calculate Eq.(\ref{Z1}) by closing
the contour of integration at infinity (which is correct for $n <
2 M - N$) and using the Cauchy formula of pole integration:
\begin{eqnarray}
\label{Z2}
Z_{n,N}(\epsilon)=(2\pi)^N (C_M)^N\sum_{\{{k_i}\}}
\frac{\prod _{i=1}^N (\epsilon-i\mu _{k_i})^n}{\prod _{i=1}^N
\prod _{j=1}^{2 M-N} (\mu_{k_i} - \mu _{p_j})},
\end{eqnarray}
where the sum is taken over all possible  subsets $ \{ k_1,\dots
,k_N \}$ of the set of indices $k\in \{ 1, \dots ,M \}$ of the
poles $\mu_{k}+iE_{i}$ on the upper half plane and
$\{ p_1,\dots ,p_{2 M-N} \} \equiv \{
1,\dots ,M,M+1, \dots 2M \} \setminus \{ k_1,\dots ,k_N \}$
includes the numeration of poles $\mu_{k}-iE_{i}$ with the
condition $\mu_{M+j}=-\mu_j, \; j=1, \dots M$. This formula will
be a starting point of the further analyzes.

\subsection{Color--flavor transformation}

%{ \it \large Color--flavor transformation.}---
Now we define another, dual representation of the partition function
in which the replica space and the space of matrix indices interchange.
Usually (see e.g. \cite{KMez}) such a representation is obtained from the
fermionic replica
sigma-model derived using the Hubbard-Stratonovich transformation.

In our case the potential (\ref{Pot}) is
not Gaussian and the  Hubbard-Stratonovich transformation is not
useful. But nevertheless it appears that
one can make a transformation of the partition function given by
an integral over original $N \times N$ Hermitian
matrices (following \cite{Zirn} let us call the space where
they are acting a color space)
and pass to the integration over
matrices, which are acting in replica space (let us call it flavor
space).
Similar type of transformations effective in
supersymmetric $\sigma$-models~\cite{Zirn} are called color-flavor
transformations.

In order to make the color-flavor transformation let us transform
the determinants in the integrand of Eq. (\ref{Z1}) into the
Gaussian integral over superfields $\Phi_i^a=(\{\Psi
_i^{\alpha}\};\{\phi_i^{(1),k}\};\{\phi_i^{(2),p}\})$ with
$a = \{\alpha;k;p\}\equiv \{1\dots n;1\dots {M };\\ 1\dots {M }\}$,
where n is the number of fermionic components $\Psi _i^{\alpha}$
while M is the number of bosonic ones $\phi_i^{(1),k},
\phi_i^{(2),p}$. The supersymmetric structure of the fields
$\Phi_i^a $ which can be regarded as $i=1,\dots N$ vectors in
(n+2M) dimensional flavor superspace, is dictated by the fact that
the partition function Eq.(\ref{Z1}) is a product of a certain
number of determinants and inverse determinants.
Using these superfields we obtain

\begin{eqnarray}
\label{Z3} Z_{n,N}(\epsilon)=\frac{(i^{-n} C_M)^N}{(2 \pi)^{2 M N}}
\int d H \prod_{i,a}d\Phi_i^{+a} d\Phi _i^a \exp(-\Phi_i^{+a}
\tilde {V}_{ij}^{ab} \Phi _j^b).
\end{eqnarray}
The indices $i,j=1\dots N$ correspond to the original color space
while indices $a$ and $b$ are forming a basis in the flavor space.
The matrix $\tilde {V}_{ij}^{ab}= diag \{[i \epsilon\delta _{ij}-i
H_{ij}] \delta_{\alpha\beta} ; [\mu_k\delta_{ij}+ i
H_{ij}]\delta_{kp} ; [\mu_k\delta_{ij}- i H_{ij}]\delta_{kp}\}$ is
diagonal
%%\begin{eqnarray}
%%\label{V1}
%%\left(
%%\begin{array}{ccc}
%%(i\epsilon \delta _{ij}-H_{ij})\delta _{\alpha\beta} & 0 & 0 \\
%%0 & 0 & (\mu _k \delta _{ij}+i H_{ij})\delta_{kp} \\
%%0 & (\mu _k \delta _{ij}-i H_{ij})\delta_{kp} & 0 \\
%%\end{array}
%%\right),\nonumber
%%\end{eqnarray}
with $\alpha ,\beta =1\dots n$; $k,p=1\dots {M}$ and Kronecker
$\delta$-s. As one can see the integral over Hermitian random
matrices $H$ in the expression (\ref{Z3}) simply produces a Dirac
delta-function $(2 \pi)^{N(N+1)/2}\prod_{i,j}\delta
[{\Phi}_i^{+a}g^{ab}\Phi _j^b]$ with
$g^{ab}=diag\{\delta_{ij},\delta_{kp},-\delta_{kp}\}$.

%%%%%%%%%%%%%%%%%%%%%%%%%%%%%%%%%%%%%%%%%%%%%%%%%%%%%%%%%%%%%%%%%%%%%%%%
%%%%%%%%%%%%%%%%%%%%%%%%%%%%%%%%%%%%%%%%%%%%%%%%%%%%%%%%%%%%%%%%%%%%%%%
For further treatment it is convenient to make use of the following
trick: we introduce a set of real numbers ${\omega
_1,\dots,\omega _N}$ and instead of the delta-function
$\prod_{i,j}\delta
[{\Phi}_i^{+a}g^{ab}\Phi _j^b]$ we
substitute
$\lim_{\{\omega_i\}\rightarrow 0}\prod_{i,j}\delta
[{\Phi}_i^{+a}g^{ab}\Phi _j^b -\omega_i\omega_j \delta_{ij}]$.
Then after the rescaling of the superfields
$\Phi_i^a\rightarrow\omega _i\Phi _i^a$ one  obtains
\begin{eqnarray}
\label{Z4} Z_{n,N}(\epsilon)&=&\frac{(i^{-n} C_M)^N}{(2 \pi)^{2 M N-N(N+1)/2}}\lim
_{\omega _i\rightarrow 0}\prod _{i,j,k=1}^N \prod_{a}\omega
_{i}^{2(2M-n-N)}\int
d {\Phi}_k^{+a} d \Phi _k^a  \nonumber\\
&\cdot&\prod_{i,j}\delta [{\Phi}_i^{+a}g^{ab}\Phi
_j^b-\delta_{ij}] \exp ({- \omega_i^2}{\Phi}_i^{+a}g^{ab}{V}_{ij}^{bc}
\Phi _j^c),
\end{eqnarray}
with
\begin{eqnarray}
\label{V2} {V}_{ij}^{ab}=
\left(%
\begin{array}{ccc}
  i \epsilon \delta_{ij}\delta_{\alpha \beta} & 0 & 0 \\
  0 & \mu_k \delta_{ij}\delta_{kp} & 0 \\
 0 & 0 & - \mu_k \delta_{ij}\delta_{kp}\\
\end{array}%
\right)
\end{eqnarray}
and the summation is understood over repeating indices $b$.
The argument of the new $\delta$- function in Eq.(\ref{Z4}) imposes
that all possible field configurations contributing to the
integral are $N$ {\em normalized} vectors in the $(n+2M)$
dimensional complex superspace with metric $g^{ab}$. By definition
this space can be called complex Stiefel manifold and is equivalent to the
homogeneous factor space $S={SU(n\mid M,M)\over{SU(n\mid M-N,M)}}$
of pseudounitary supergroups. Therefore  the formula
(\ref{Z4}) is nothing but a replica $\sigma$ model formulation of
the model  (\ref{Z1}) in the form of graded IZ type integral over
Stiefel manifold in the $(n+2M)$ dimensional flavor superspace:
%%%%%%%%%%%%%%%%%%%%%%%%%%%%%%%%%%%%%%%%%%%%%%%%%%%%%%%%%%%%%%%%%%%%%%%%%
%%%%%%%%%%%%%%%%%%%%%%%%%%%%%%%%%%%%%%%%%%%%%%%%%%%%%%%%%%%%%%%%%%%%%%%%
\begin{eqnarray}
\label{Z5} Z_{n,N}(\epsilon)= \frac{(i^{-n} C_M)^N}{(2 \pi)^{2 M
N-N(N+1)/2}}\lim _{\omega _i\rightarrow 0}\prod _{i=1}^N\omega
_{i}^{2(2M-n-N)} \int_{U\in {\cal S}} d U \exp \{- Str(\Omega
U^{-1}VU)\},
\end{eqnarray}
where $U^{-1}= g U^{+} g$, the matrices
$V=diag\{i \epsilon _1\dots i \epsilon _n, \mu _1\dots \mu_{M},
-\mu _1\dots -\mu _{M}\};$
%is the diagonalization of $V^{ab}$,
%%in the limit $\epsilon _1,\dots ,\epsilon _n \rightarrow \epsilon$ and
$\Omega =diag\{ \underbrace{0\dots 0} _{n},\omega _1^2\dots \omega
_N^2,\underbrace{0\dots 0} _{2M-N} \}$ and the limit $\epsilon
_1,\dots ,\epsilon _n \rightarrow \epsilon$ is supposed.

The evaluation of standard (super) IZ integral over the group
manifold is based on the analyze of heat kernel equation in the
Hermitian matrix space and on calculation of induced metric on the
subspace of diagonal matrices \cite{IZ}, \cite{Guhr}, \cite{Alfaro}. In our
case, since the tangent to the Stiefel manifold vector defined by
$d H =UdU^{-1} \setminus GdG^{-1}$, with $U\in SU(n\mid M,M)$, $G\in
SU(n\mid M- N,M)$, the corresponding induced metric does not
coincide with the Vandermonde determinant, but with some
modification.

Another elegant calculation of IZ integral over unitary supergroups
has been made in \cite{Szabo} by use of Duistermaat-Heckman
theorem. In \cite{Fyo} it has been shown, that the theorem can be applied
to pseudounitary supergroups as well.
In Appendix A we apply this theorem to our IZ integral
over manifold $SU(n\mid M,M)/SU(n\mid M-N,M)$. Leaving the
mathematical details of calculations aside we present here only
the answer. Let us introduce a family of maps $\sigma: \{1,2,\dots
,N\}\rightarrow \{1,\dots ,M\}$ of the indices of the original
color space into the subset of indices of the positive root part
of the flavor superspace. Correspondingly $\bar \sigma$ will
denote the complement part of the $\sigma$: $\bar \sigma =
\{1,\dots ,2M\} \setminus \{\sigma (1), \dots ,\sigma (N)\}$.  Then
the answer is (see Appendix A for details)
\begin{eqnarray}
\label{IZ} \int _{U\in {\cal S}} d U \exp \{-Str(\Omega U^{-1}V
U)\}=(2 \pi)^{2MN-N(N-1)/2} \sum _{\{\sigma \}}\frac{\det
[\exp {-\omega _k^2\mu _{\sigma (p)} }]} {\Delta _\sigma
(\Omega)\Delta _\sigma (V)}
\end{eqnarray}
where $\alpha ,\beta =\overline {1,n}$; $k,p= \overline {1,N}$ are
fermionic and bosonic components correspondingly,
the sum is taken over all possible maps $\sigma$ and
$\Delta _\sigma (\omega)$, $\Delta _\sigma (V)$ are defined by
the induced metric as
\begin{eqnarray}
\label{Jacobian}
\Delta_{\sigma} (\nu)=\frac{\prod _{i=1}^N \prod _{p=1}^{2 M-N}
(\nu _{\sigma (i)} - \nu _{\bar \sigma (p)})
(\nu _{\sigma (i)}-\nu _{\sigma (j)})}
{\prod _{\alpha =1}^{n}\prod _{i=1}^N(\nu _{\alpha}-\nu _{\sigma (i)})}.
\end{eqnarray}
The formulae~(\ref{IZ}) and~(\ref{Jacobian}) present the extension
of IZ matrix integral for the case of integration over Stiefel
manifold $\cal{S}$. It is clear, that the expressions (\ref{IZ})
(\ref{Jacobian}) were obtained for integer values of $n$. Now
taking the limit $\omega_i\rightarrow 0$ in (\ref{Z5}) we will
reproduce the same formula (\ref{Z2}) for replica partition
function as it was obtained in original color formulation. This
demonstrates the equivalence of the color and flavor formulations
of the model under consideration.

\section{ Toda Lattice Equations}
\noindent
In the present section we derive Toda lattice equations in both, color and flavor spaces
respectively. The first part is devoted to the so called $\tau$ function,
defined for a unitary ensemble with any given distribution function, and TLE
for it in color space with initial conditions corresponding to the RMT Eq. (\ref{Pot}).
In the second part we show how the fermionic and bosonic flavors involved in
the replica partition functions $Z_{n,N}$, connect them with integrable hierarchies.

\subsection{TLE in color space}
%{\it \large TLE in color space.}---
It is easy to see, that the replica
partition function (\ref{Z1}) is the $T\rightarrow 0$ limit of so
called $\tau$-function $\tau_N(T,\epsilon)=\int d H \exp Tr[T H+\tilde
{V}(H)]$ of the model with the potential $\tilde{V}(H)=n\ln
(\epsilon-H)+V(H)$, which can be regarded also as a matrix
Fourier transform. $N \times N$ matrix $T$, being conjugate
variable to energy matrix $H$, can be regarded as an effective time.
 The $\tau$-function can be calculated exactly as
\cite{Morozov}
\begin{eqnarray}
\label{tau}
\tau _N(T;\epsilon)= {1\over {\Delta(T)}}\det [{W_\delta}(t_j)]_{\delta
,j=0,\dots ,N-1},
\end{eqnarray}
where $\Delta (T)$ is the Vandermonde determinant corresponding to
the matrix $T$ and
\begin{eqnarray}
{W_\delta}(t)=\int d \lambda \lambda ^\delta
\exp\{\lambda t + \tilde{V}(\lambda)\}
\end{eqnarray}
with $\delta =0,\dots ,N-1$.
In the $t_i\rightarrow 0$ limit (\ref{tau}) gives us the
replicated partition function $Z_{n,N}(\epsilon)=\tau_N(0,\epsilon)$ and
$\tau_N(t,\epsilon)=\tau_N(T,\epsilon)\mid_{\{t_i\}=t}$ can be expressed via
Hankel determinant
\begin{eqnarray}
\label{pf} \tau_N(t,\epsilon)=\det_{\delta,j=0}^{N-1}[\partial
^{j+\delta}_t W_0(t)].
\end{eqnarray}
Then it is clear \cite{Kanz1},\cite{Morozov}, that $\tau_N(t,\epsilon)$ will
satisfy the TLE
\begin{eqnarray}
\label{TLE} \tau_N(t,\epsilon)\partial_t^2\tau_N(t,\epsilon)-(\partial_t
\tau_N(t,\epsilon))^2
=\tau_{N+1}(t,\epsilon)\tau_{N-1}(t,\epsilon)\nonumber
\end{eqnarray}
with the initial condition

\begin{eqnarray}
\tau_1(0,\epsilon)=2 \pi C(M) \sum _{k=1}^{2M}[{(\epsilon -i \mu
_k)^n}/{\prod _{p=1;p\neq k}^{2M} (\mu _k-\mu _p)} ].
\end{eqnarray}

\subsection{TLE in flavor (supersymmetric) space}
%{\it \large TLE in flavor (supersymmetric) space.}---
After some
algebraic manipulations the replicated partition function
(\ref{Z2}) can be rewritten as
\begin{eqnarray}
\label{Z11} &&Z_{n,N}(\epsilon)={(C_M)^N \over {\cal V} _{2M}(\mu
_1\dots\mu _{2M})}\lim _{\epsilon _\alpha\rightarrow\epsilon}
{1\over {\cal V} _{n}(\epsilon _1\dots\epsilon _n)} \\
&&\sum_{\{\sigma\}}(-1)^{\pi (\sigma)} {\cal V} _{N+n}(\epsilon
_1\dots\epsilon _n,\mu _{\sigma (1)}\dots\mu _{\sigma (N)})
{\cal V}_{2M-N}(\mu _{\bar \sigma (1)}\dots\mu _{\bar \sigma
(2M-N)})\nonumber
\end{eqnarray}
where  ${\cal V}(\{\dots \})$ is the Vandermonde determinant
defined by the set of arguments $\{ \dots \}$ and $\pi (\sigma)$
is the parity of permutation of $1,\dots ,N,N+1,\dots ,{M}$ to
$\sigma (1),\dots ,\sigma (N), \bar \sigma (1)\dots ,\bar \sigma
({M}-N)$. This form of $Z_{n,N}(\epsilon)$ can be understood as a
formula for a determinant calculated by minors
\begin{eqnarray}
\label{Z12} Z_{n,N}(\epsilon)={(i^{-n}C_M)^N \over {\cal V} _{2M}(\mu
_1\dots\mu_{2M})}\lim _{\epsilon _\alpha\rightarrow\epsilon}
{I^{-1}\det [A]\over {\cal V} _{n}(\epsilon _1\dots\epsilon _n)}
\end{eqnarray}
where we have introduced the $(n+2M)\times (n+2M)$matrix $A$ as
\begin{eqnarray}
\label{det}
A=\left (
\begin{array}{ccccccc}
I_1 & \delta_1I_1 &  \dots & \delta_1^{n+N-1} I_1 & 0 & \dots & 0\\
I_2 & \delta_2I_2 &  \dots & \delta_2^{n+N-1} I_2 & 0 & \dots & 0\\
\vdots & \vdots &  \vdots & \vdots &  \vdots & \vdots & \vdots \\
I_n & \delta_nI_n & \dots & \delta_n^{n+N-1} I_n & 0 &  \dots & 0\\
\bar{I}_1 & \bar \delta_1 \bar{I}_1 & \dots & \bar
\delta_1^{n+N-1} \bar{I}_1 &  \bar{I}_1  &
\dots &\bar \delta_1^{2M-N-1} \bar{I}_1\\
\vdots & \vdots & \vdots & \vdots &  \vdots & \vdots & \vdots \\
\bar{I}_M & \bar \delta_{M} \bar{I}_M &  \dots & \bar
\delta_{M}^{n+N-1} \bar{I}_M &  \bar{I}_M
& \dots &\bar \delta _{M}^{2M-N-1} \bar{I}_M \\
0 & 0 &  \dots & 0 &  \bar{I}_{M+1}
& \dots &  \bar \delta _{M+1}^{2M-N-1} \bar{I}_{M+1} \\
\vdots & \vdots & \vdots  &  \vdots & \vdots &  \vdots & \vdots \\
0 & 0 &  \dots & 0 &  \bar{I}_{2M} & \dots
&  \bar \delta _{2M}^{2M-N-1} \bar{I}_{2M} \\
\end{array}
\right),
\end{eqnarray}
involving the functions $I_{\alpha}=\exp (i \epsilon _{\alpha}),\;
\alpha =1, \dots n;\;
\bar{I}_r=\exp\mu _r,\;r=1,\dots 2M;\;
I=\prod_{\alpha,r=1}^{n,2M} I_{\alpha}\bar I_r $ and the notations
$\delta _{\alpha}=\epsilon _{\alpha} {\partial \over \partial
\epsilon _{\alpha}}; \;
\bar \delta _r = \mu _r {\partial \over
\partial \mu _r}.$ Define
$\delta^{1+2M}=\epsilon {\partial \over \partial \epsilon }
+\sum_{r=1}^{2M}(\bar \delta_r)$.

Now, following the technique developed in \cite{SV2},
the modified graded Toda lattice equation for
replicated partition function $Z_{n,N}(\epsilon)$ will
follow directly from its expression (\ref{Z12}) as a determinant and
two sets of Sylvester
identity \cite{Syl} for the matrix $A$ (see Appendix B for a details).
As a result, the following equation is
established
\begin{eqnarray}
\label{GTLE} n-i \partial _{\epsilon}\delta ^{1+2M}\ln
[Z_{n,N}(\epsilon)]&=& \nonumber \\
n\frac{Z_{n+1,N}(\epsilon)Z_{n-1,N}
(\epsilon)}{Z_{n,N}^2(\epsilon)}&+&n\frac{Z_{n+1,N-1}
(\epsilon)Z_{n-1,N+1}(\epsilon)}{Z_{n,N}^2(\epsilon)},
\end{eqnarray}
where $n < 2 M-N$, when $Z_{n,N}$ is finite. It has a graded form
since contains derivatives over both, fermionic and bosonic
parameters. One can see here, that in the replica limit
$n\rightarrow 0$ the right hand side (RHS) of this equation
involves a factorized product of fermionic and bosonic partition
functions $Z_{n=1}$ and $Z_{n=-1}$ correspondingly. Moreover, the
second term in RHS shows that indeed fermionic, bosonic and
supersymmetric partition functions for any N belongs to the same
integrable hierarchy.

In the limit $N\rightarrow \infty$ (but $N < M$),
when $Z_{n,N-1}=Z_{n,N+1}=Z_{n,\infty}$, the equation (\ref{GTLE})
simplifies to
\begin{eqnarray}
\label{GTLE1} n-i \partial_{\epsilon}\delta^{1+\infty}\ln
[Z_{n,\infty}(\epsilon)]= 2
n\frac{Z_{n+1,\infty}(\epsilon)Z_{n-1,\infty}
(\epsilon)}{Z_{n,\infty}^2(\epsilon)},
\end{eqnarray}
where $\delta^{1+\infty}=\epsilon {\partial \over \partial \epsilon }
+\sum_{r=1}^{\infty}(\bar \delta_r)$.

Similar to (\ref{GTLE1}) type of equations, but for a finite amount
of mass parameters $\mu$ were considered earlier in \cite{SV3},
\cite{SV2}, \cite{Akemann}, \cite{Mironov}. The factorization
properties of Toda lattice equations in the replica limit have been
analyzed in \cite{SV3}.

As we see the equation (\ref{GTLE}) differs essentially from the
known equation (\ref{GTLE1}), which defines integrable Toda
lattice model \cite{Toda}. Since the equation (\ref{GTLE}) appears
for the exactly calculable partition function (\ref{Z2}) of the
unitary matrix model with potential (\ref{Pot}), one can
conjecture the existence of corresponding integrable lattice chain
model. It is definitely interesting to investigate the interplay
between color and flavor degrees of freedom in the equation
(\ref{GTLE}) further and analyze its consequences.

%%%%%%%%%%%%%%%%%%%%%%%%%%%%%%%%%%%%%%%%%%%%%%%%%%%%%%%%%%%%%%%%%%%%%
\section{Density of States, Mean Conductance and Two Point Correlation Function }
\noindent

%{\it The two point correlation function.}---
In the original
articles \cite{Muttalib}, \cite{Ismail} where the model with the
potential (\ref{Pot}) was initiated,
%Muttalib {\it et al}
the authors, by use
of q-Hermite orthogonal polynomials, have analytically calculated
the $N\rightarrow\infty$ asymptotic of the two point correlation
function. For the appropriately renormalized two point kernel in
the bulk of the spectrum exact expression approximately becomes
(except very close to the origin)
\begin{equation}
\label{K2}
K(\xi,\eta)\simeq\frac{\beta}{2 \pi}\frac{\sin[\pi(\xi-\eta)]}
{\sinh[(\xi-\eta)\beta/2]},
\end{equation}
which is translational invariant. In its $\beta\rightarrow 0$
limit $K(\xi,\eta)$ reproduces the universal behavior of
Wigner-Dyson RMT,
%%GUE $\sin[\pi(\xi-\eta)]/\pi(\xi-\eta)$,
while for the large $\beta$,
as authors of  \cite{Muttalib} have argued, the model approaches
to the uncorrelated systems with Poisson  distribution.

In our approach the calculation of the density of states $\rho(x)$ 
%$\langle Tr\frac{1}{(x-H)}\rangle _H$
will be reduced to Cauchy integration over poles of the distribution
function in the
\begin{eqnarray}
\label{dist}
\rho(x) = \frac{1}{ Z_{0,N}(\epsilon)}\int_{-\infty}^{\infty} dH tr\delta(x-H) P(H) =\frac{N (C_M)^{N}}{Z_{0,N}(\epsilon)}
\int \prod_{i=2}^{N}dx_i \prod_{i,j =1}^{N}(x_i-x_j) \prod _{k,j=1}^{M,N}
\frac{1}{(\mu _k + i x_j)(\mu _k - i x_j)},
\end{eqnarray}
where the normalization factor  $ Z_{0,N}(\epsilon)$ is defined by the expression
(\ref{Z2}), for the case $n=0$. In order to analyze the large $N$
limit we will consider  $N=M \rightarrow \infty$. In this case the calculation
of the Cauchy integrals over poles in our problem essentially simplifies.
From (\ref{Z2}) one can deduce 
\begin{eqnarray}
\label{Z0} 
Z_{0,M}(\epsilon)=(2\pi)^M (C_M)^M M!
\prod _{k,p=1}^M \frac{1}{ (\mu_{k_i} +\mu _{p_j})}.
\end{eqnarray}

Now, in order to take the Cauchy integral over poles in (\ref{dist})
we close the integration contours at infinity of upper half plane
and notice, that because of presence of Vandermonde determinant,
the integration variabes $x_i, \; i=2, \dots, M$ will utilize 
$M-1$ poles, all different. Only one pole will be left untouched
and we will have the summation over it in the result:
\begin{eqnarray}
\label{rho2}
\rho(x)=\frac{1}{2 \pi} \sum_{k=1}^M \frac{2 \mu_k}{x^2+ \mu_k^2}
\prod_{q \neq k}^{M}\frac{(x-i \mu_q)(\mu_k+\mu_q)}
{(x+i \mu_q)(\mu_k-\mu_q)}=\frac{1}{ \pi} \sum_{k=1}^M \frac{ \mu_k}{x^2+ \mu_k^2}.
\end{eqnarray}
Constituents of this formula have the following meaning. The factor
$(x^2+ \mu_k^2)\prod_{q\neq k}^M(x+i  \mu_q)$
in the denominator comes from the distribution function of the remnant eigenvalue
$x$, the part of which, namely  $\prod_{q\neq k}^M(x-i  \mu_q)$, was cancelled
with the similar bracket in the Vandermonde determinant. The factor $\prod_{q\neq k}^M(x-i  \mu_q)$
in the nominator is the remnant from the Vandermonde determinant. The factor
$\prod_{q\neq k}^M \frac{(\mu_k+\mu_q)}{(\mu_k-\mu_q)}$ is left after mutual cancellations
between the residues of integrated poles, the Vandermonde determinant and the normalization factor
$Z_{0,N}(\epsilon) $ defined by formula (\ref{Z0}). The second part of the
formula (\ref{rho2}) is an identity.

Hence, we have obtained a simple formula:
\begin{eqnarray}
\label{rho}
\rho(x)=\frac{1}{\pi} \sum_{k=1}^M Im(\frac{1}{x-i
\mu_k}).
\end{eqnarray}
One will arrive at the same formula (\ref{rho}) taking a replica limit for the one
point Green's function
$G(x)=\lim _{n\rightarrow 0}n^{-1}\partial Z_{n,M}(x)/\partial x$ and calculating the DOS.

The model under consideration with potential (\ref{Pot}) first was 
considered in \cite{Muttalib},
in order to describe the disordered conductors  
connecting matrices $H$ with the ensemble of random transfer matrices $T$ as
\begin{eqnarray}
\label{tm}
H^2= \frac{1}{4}(T T^+ + T^+ T -2 I).
\end{eqnarray}

Therefore substituting $x=\sinh[\beta \chi/2]$
($e^{\beta \chi/2}$  is the eigenvalue of $T$) into the $\rho(x)$ one will obtain
an exact expression for the density of  $\chi$:
\begin{eqnarray}
\label{cr}
\rho(\chi)=\frac{dx}{d\chi}\rho(x).
\end{eqnarray}
which exhibits crystallization at the integer values $n$, as it is presented 
on Fig.1 for two values of $\beta$. This phenomenon for different ensembles
was observed earlier in the articles \cite{Bogomol}, \cite{BR}, \cite{Lama}.

\begin{figure}
\includegraphics{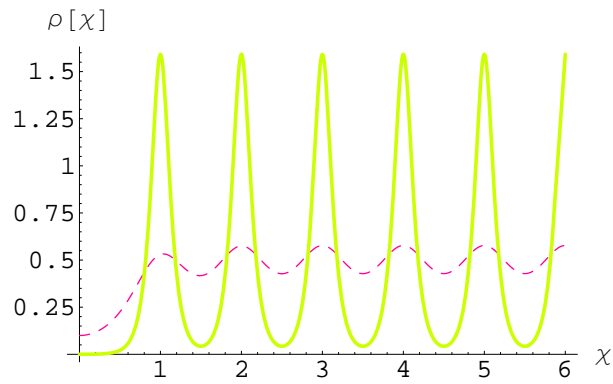}
\caption{$\chi$ eigenvalue densities at $\beta=6$ (dashed line) and $\beta=20$ 
(solid line).
The connection with transmission eigenvalues of $\cal T$ is given by ${\cal T} =
 \cosh[\beta \chi /2]^{-2}$ }
\label{Fig2}
\end{figure}

The  eigenvalues $x_i$ of $H$ and $\chi_i$ are connected with the conductance 
$g=tr(tt^+)/2$ ($t$ is the transmission matrix) as
\begin{eqnarray}
\label{con}
g=\sum_{i=1}^{N} \frac{1}{1+ x_i^2}= \sum_{i=1}^{N} \frac{1}{\cosh[\beta \chi _i/2]^2}
\end{eqnarray}
and by the same type of Cauchy integration over poles as in $\rho(x)$
one can find the average conductance
\begin{eqnarray}
\label{con1}
\langle g \rangle = \sum_{n=1}^M \frac{1}{1+\mu_n^2}= \sum_{n=1}^M \frac{1}{1+\cosh[n \beta/2]^2}.
\end{eqnarray}

In the same way one can calculate the two point
correlation function $R_2[x,y]$, again in the limit $N=M \rightarrow \infty$
\begin{eqnarray}
\label{R20}
R_2(x,y) = \frac{1}{ Z_{0,N}(\epsilon)}\int_{-\infty}^{\infty} dH 
tr\delta(x-H)tr\delta(y-H) P(H).
\end{eqnarray}
By Cauchy integration over poles, in this case, we should now leave untouched 
two poles of the distribution
function $P(H)$ (see formula (\ref{dist})). Therefore, in the result we will 
have a sum over all
possible positions $m$ and $n$ of that poles. The answer reads:
 \begin{eqnarray}
\label{R2}
R_2(x,y)=\frac{1}{\pi^2}\sum_{m,n}^M\frac{\mu_m \mu_n }
{(x^2+\mu_n^2)(x^2+\mu_m^2)(y^2+\mu_n^2)(y^2+\mu_m^2)}\nonumber\\
\times\prod_{k\neq m,n}\frac{(x-i
\mu_k)(\mu_m+\mu_k)(y-i\mu_k)(\mu_n+\mu_k)} {(x+i
\mu_k)(\mu_m-\mu_k)(y+i \mu_k)(\mu_n-\mu_k)}. \; \;\; \;\;\quad
\end{eqnarray}
Details of the calculation of $R_2(x,y)$ repeat the detales of the one for 
the density of states $\rho(x)$,
presented above. Since here we hold two eigenvalues $x$ and $y$ untouched,
the formula (\ref{R2}) is similar (but not equal) to the product of two 
structures (\ref{rho2}) of
$\rho(x)$ with corresponding adjustment of the summation procedure.
Unlike to the case of $\rho(x)$,  I was not able to find a simplification
of the final formula (\ref{R2}), as it was done in the second part of 
Eq.(\ref{rho2}).

In spite of explicit presence of the imaginary unite $i$ in the
expression (\ref{R2}) the resultant $R_2(x,y)$ is real. After unfolding,
namely after the re-scaling $x\rightarrow \xi$, defined by
$d\xi/dx = \rho(x)$, $R_2(\xi,\eta)$ is
 pictured in Fig.2 by dots for
$\beta=0.3$. Moreover, the numerical analyze of the expression
(\ref{R2}) after unfolding shows that $R_2(\xi,\eta)$ in fact is a
function of difference $\xi-\eta$.

Numerical analyses show, that consideration of other potentials, 
say by taking the product in
formula (\ref{Z1}) from $k=0$ or $k=2$, after appropriate
unfolding gives us the same behavior as approximately defined by
formula (\ref{K2}). 
%This confirms the statement by Bogomolny {\it
%et al.} in \cite{Bogomol}, that a smooth change of the potential
%of the model does not affect the universality class.

%\begin{figure}
%\includegraphics{rho.eps}
%\caption{Transmission eigenvalue densities at $\beta=6$ (dashed
%line) and $\beta=20$ (solid line). }\label{Fig2}
%\end{figure}
\begin{figure}
\includegraphics{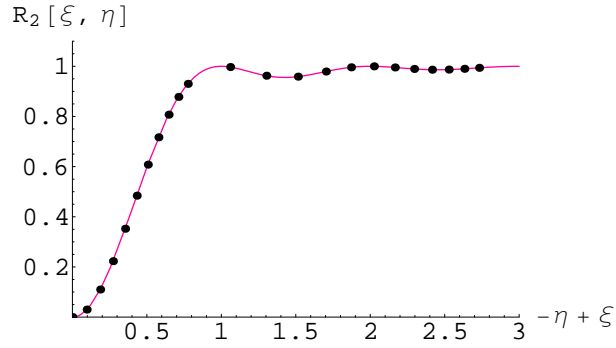}
\caption{The two point correlation function $R_2[\xi-\eta]$ calculated
exactly by Cauchy integration over poles and after unfolding
with $\rho(x)$ (dots). The solid line represents $R_2[\xi-\eta]$
obtained from the formula (\ref{K2}), both at $\beta=0.3$.}\label{Fig1}
\end{figure}

%\begin{eqnarray}
%\label{R2} R_2(x,y)=\frac{1}{\pi^2}\sum_{m,n=1}^M\frac{\mu_m \mu_n (x-y)^2
%(\mu_n+\mu_m)^2}
%{(x^2+\mu_n^2)(x^2+\mu_m^2)(y^2+\mu_n^2)(y^2+\mu_m^2)}\nonumber\\
%\times\prod_{k\neq m,n}\frac{(x-i
%\mu_k)(\mu_m+\mu_k)(y-i\mu_k)(\mu_n+\mu_k)} {(x+i
%\mu_k)(\mu_m-\mu_k)(y+i \mu_k)(\mu_n-\mu_k)}\; \;\; \;\;\quad
%\end{eqnarray}
%%\begin{equation}\label{rho}
%%\rho(\chi)= \frac{\beta}{2\pi}\cosh[\frac{\beta
%%\chi}{2}]\sum_{k=1}^M Im\left[\frac{1}{\sinh[\beta \chi/2]-i
%%\cosh[n \beta/2]}\right],\nonumber
%%\end{equation}

\section*{Conclusions}
\noindent

We give a sigma-model formulation of the RMT with confining
potential (\ref{Pot}). The replica partition function of the model
has been calculated exactly in both, original color space and in
flavor space (after a color-flavor transformation) with the same
result. In both spaces we have established connections between the
RMT Eq.(\ref{Pot}) and exactly solvable lattice equations. A
generalization of graded TLE is defined and solved.

We have exactly calculated the probability density and the two
point correlation function of the model at $N\rightarrow\infty$
and confirmed the known results obtained earlier. The effect of
crystallization of transmission eigenvalues has been observed.

\section*{Acknowledgments}
\noindent

I am greatly thankful to V. E. Kravtsov for numerous stimulating
discussions and critical reading of the paper. I would like also
to acknowledge G. Akemann, E. Bogomolny, B. N.
Narozhny, J. J. M. Verbaarschot and especially E. Kanzieper for comments.

%%%%%%%%%%%%%%%%%%%%%%%%%%%%%%%%%%%%%%%%%%%%%%%%%%%%%%%%%%%%%%%%%%%%%%%%%%
%%%%%%%%%%%%%%%%%%%%%%%%%%%%%%%%%%%%%%%%%%%%%%%%%%%%%%%%%%%%%%%%%%%%%%%%%%%
\section*{ Appendix A. Evaluation of Supersymmetric
Itzikson-Zuber Integral over $SU(n \mid M,M)/SU(n \mid M-N,M)$ }
\noindent

In this appendix we present a simple derivation of the supersymmetric
extension of the Itzikson-Zuber formula over Stiefel manifold
$SU(n\mid M,M)/SU(n\mid M-N,M)$.
\begin{eqnarray}
\label{IZA} {\cal I}(X,Y; r)= \int_{U \in SU(n \mid M,M)/SU(n\mid
M-N,M)} d U \exp \{r \; Str(X U^{-1} Y U)\}
\end{eqnarray}

In this calculation one may follow the standard technique
\cite{IZ} developed in \cite{Guhr}, \cite{Alfaro} for unitary supergroups
based on the solution of the heat kernel equation. In \cite{And} it was
argued, that the heat kernel equation technique can be applied to pseudounitary
supergroups. However we will present here the
calculations based on the application of Duistermaat-Heckman theorem
to integrals over supermanifolds, which states that under certain
conditions the quasi-classical saddle-point approximation is exact
\cite{ASw}. This theorem was successfully applied to IZ integral
over unitary supergroups $U(n\mid m)$ in \cite{Szabo} and pseudounitary
supergroups $U(n\mid m,k)$ in \cite{Fyo}.
In what follows we will assume, that this criteria are met
by our integral (\ref{B1}) as well and evaluate the integral in the
saddle-point approximation. The fact, that final result reproduces the
same answer (\ref{Z3}) for $Z_{n,N}(\epsilon)$, as we got by simple calculation
of the integral (\ref{Z2}) by the Cauchy formula confirms, that
this assumption is correct.
The same answer one will obtain
following the standard technique \cite{Guhr}, \cite{Alfaro}.

We need to find the extremum of the function
\begin{eqnarray}
\label{ex}
    {\cal A}(U)= Str(X U^{-1} Y U),
\end{eqnarray}
where $U \in SU(n\mid M,M)/SU(n\mid M-N,M)$ and $X$ and $Y$
are diagonal matrices from the expression (\ref{Z5}). It is easy
to find the saddle-point equation
\begin{equation}\label{sp}
    [X,\; U^{-1} Y U]=0
\end{equation}
from the condition $\delta {\cal A}(U)= Str\{(X U^{-1} Y U - U^{-1}
Y U X) U^{-1}\delta U\}=0$.

Since we have $X =diag\{ \underbrace{0\dots 0} _{n},x_1\dots
x_N,\underbrace{0\dots 0} _{2M-N} \} $ and \\
$U \in SU(n \mid M,M)/SU(n\mid M-N,M)$ the only
solutions  $U$ of (\ref{sp}) (up
to irrelevant $U \in SU(n \mid M-N,M)$) are those matrices, which
permute the diagonal elements of the matrix $Y$ in three- fermionic,
positive and negative metric parts of the bosonic sectors separately.
In other words
the set of solutions of (\ref{sp}) is
\begin{eqnarray}
\label{sol} U_{a b}^{(0)}= \left(%
\begin{array}{ccc}
 \bar{\prod}_{\alpha \beta}& 0&0  \\
 0 & \prod_{kp}&0 \\
 0&0& \prod_{k^{\prime}p^{\prime}}^{\prime}\\
\end{array}%
\right),\qquad
\begin{array}{c}
\alpha,\beta = 1,\dots ,n\\
k,p = 1,\dots ,M\\
k^{\prime},p^{\prime} = 1, \dots ,M
\end{array}
\end{eqnarray}
where $\bar{\prod}_{\alpha \beta}$ and $\prod_{kp}
(\prod_{k^{\prime}p^{\prime}}^{\prime})$ are the elements of
the permutation groups $S_n$ and $S_{M}$ of the fermionic and
bosonic sectors respectively. Notice, that the set of permutation
elements $\prod_{ij}, \; i,j = 1, \dots N$ coincides with the set
of maps $\sigma: \{1, \dots N\} \rightarrow \{ 1, \dots M\}$
defined earlier.

We now set $U=U^{(0)}e^{i L}$ in (\ref{ex}), with
an infinitesimal Hermitian supermatrix
$L = U^{-1}\delta U \in su(n\mid M,M) \setminus su(n\mid M-N,M)$ and
expand the function
${\cal A}(U)$ up to the quadratic order in $L$. In this factor space
$L$ has only following nonzero matrix elements
$L_{ij}, \;i,j=1,\dots N; L_{ik}, \;i=1,\dots N; k=1, \dots 2M-N$
and $L_{i \alpha},\; i=1,\dots N;\;\alpha=1, \dots n$. In the second
order over $L$, ${\cal A}(U)$ is
\begin{eqnarray}
\label{au}
{\cal A}(U)&=& StrX U^{(0)}Y U^{(0)}+ StrX L U^{(0)}Y U^{(0)}L- Str\frac{1}{2}
[X, U^{(0)}YU^{(0)}]L^2\nonumber\\
&=&StrX U^{(0)}Y U^{(0)}+ \frac{1}{2} Str[X, L][U^{(0)}Y U^{(0)}, L].
\end{eqnarray}

Due to Duistermaat-Heckman
theorem in evaluation of IZ integral (\ref{IZA}) we should take the sum
over all extrema, therefore substituting (\ref{au}) to (\ref{IZA}) we
obtain

\begin{eqnarray}
\label{IZA1}
 {\cal I}(X,Y; r)&=& \sum_{\sigma \in S_{2 M}}\exp (r \sum_{i=1}^{N}
 x_i y_{\sigma(i)})\int\prod_{i,j=1}^{N}dL_{ij}\prod_{i,k=1}^{N,2M-N}dL_{ik}
 \prod_{i,\alpha=1}^{N,n}dL_{i\alpha}\nonumber\\
 &\times& \exp r \left[{1 \over 2} \sum_{i,j=1}^{N}\mid L_{ij}\mid^2(x_i-x_j)
 (y_{\sigma(i)}-y_{\sigma(j)})\right.\nonumber\\
 &+&\left.{1 \over 2} \sum_{i,k=1}^{N,2M-N}\mid
  L_{ik}\mid^2 x_i(y_{\sigma(i)}-y_{\bar{\sigma}(k)})\right.\nonumber\\
 &+&\left. \sum_{i,\alpha=1}^{N,n}\mid L_{i\alpha}\mid^2 x_i
 (y_{\sigma(i)}-\bar{y}_{\alpha})
 \right] ,
\end{eqnarray}
where $\bar{y}_{\alpha}$ are fermionic components of the diagonal
matrix $Y$ and $\bar{\sigma}$ is the complement to $\sigma$
permutation, defined after the formula (\ref{Z5}). Here we essentially
have used
$diag\{ \underbrace{0\dots 0} _{n},x_1\dots
x_N,\underbrace{0\dots 0} _{2M-N} \} $ form of the matrix $X$.

Evaluating now the Gaussian integrals in (\ref{IZA1}) over complex
bosonic $L_{ij},\;i \neq j ;\;
L_{i,k},\;i \neq k$ and the complex Grassmann variables $L_{i\alpha}$,
we arrive at
\begin{eqnarray}
\label{IZA2} {\cal I}(X,Y; r)&=&(2 \pi)^{2 M N -N(N-1)/2} (-r)^{N n
-2 M N+(N-1)N/2} \nonumber\\
&\times& \sum _{\{\sigma \}}\frac{\det [\exp {r x_i y_{\sigma(j)}
 }]} {\Delta _\sigma (X)\Delta _\sigma (Y)}
\end{eqnarray}
with
\begin{eqnarray}
\label{AJacobian}
\Delta_{\sigma} (\nu)=\frac{\prod _{i=1}^N \prod _{p=1}^{2 M-N}
(\nu _{\sigma (i)} - \nu _{\bar \sigma (p)})
(\nu _{\sigma (i)}-\nu _{\sigma (j)})}
{\prod _{\alpha =1}^{n}\prod _{i=1}^N(\nu _{\alpha}-\nu _{\sigma (i)})}
\qquad \qquad \nu = X,Y.
\end{eqnarray}

%%%%%%%%%%%%%%%%%%%%%%%%%%%%%%%%%%%%%%%%%%%%%%%%%%%%%%%%%%%%%%%%%%%%%
\section*{Appendix B. Graded Toda Lattice Equation}
\noindent

In this Appendix we show how the graded Toda lattice equation for
the replica partition function $Z_{n,N}(\epsilon)$ appears. First
we would like to demonstrate, that $Z_{n,N}(\epsilon)$ is a ratio
of certain determinants. The expression (\ref{Z2}) can be regarded
as $\epsilon_{\alpha} \rightarrow \epsilon ; \alpha =1,\dots n$
limit of n different fermionic eigenvalues. For the convenience
and compactness of formulas we will drop from further
transformations the constant factor $(2 \pi)^N (C_M)^N i^{nN}$ and
introduce the notation $e_{\alpha}=-i \epsilon_{\alpha}$ for a
while. Also, instead of fixed $\mu_1 \dots \mu_M, -\mu_1,\dots
-\mu_M$ set of parameters of the model, we consider here general
situation with different $\mu_k, \; k= 1, \dots 2M$. Condition
$\mu_{k+M}= -\mu_k$ can be easily imposed at the end. We have
\begin{eqnarray}
\label{B1} &&Z_{n,N}(\epsilon)=\lim_{\{e_{\alpha}\}\rightarrow
e}\sum_{\{{\sigma}\}} \frac{\prod_{\alpha=1}^n\prod _{i=1}^N
(e_{\alpha}-\mu _{\sigma(i)})} {\prod _{i=1}^N
\prod _{j=1}^{2 M-N} (\mu_{\sigma(i)} - \mu _{\bar{\sigma}(j)})}\nonumber\\
&=& \lim_{\{e_{\alpha}\}\rightarrow e}\sum_{\{{\sigma}\}}
\frac{\prod_{\alpha<\beta=1}^n(e_{\alpha}-e_{\beta})
\prod_{\alpha=1}^n\prod _{i=1}^N (e_{\alpha}-\mu _{\sigma(i)})
}
%%%%%%%%%%%%%%%%%%%%%%%%%%%%%%%%%%%%%%%%%%%%%%%%%%%%%%%%%
{\prod_{\alpha<\beta=1}^n(e_{\alpha}-e_{\beta})\prod _{i=1}^N
\prod _{j=1}^{2 M-N} (\mu_{\sigma(i)} - \mu _{\bar{\sigma}(j)})
}\nonumber\\
&\times&\frac{\prod_{i<j=1}^N(\mu_{\sigma(i)}-\mu_{\sigma(j)})
\prod_{k<p=1}^{2M-N}(\mu_{\bar{\sigma}(k)}-\mu_{\bar{\sigma}(p)})}
{\prod_{i<j=1}^N(\mu_{\sigma(i)}-\mu_{\sigma(j)})
\prod_{k<p=1}^{2M-N}(\mu_{\bar{\sigma}(k)}-\mu_{\bar{\sigma}(p)})},
\end{eqnarray}
where the nominator and denominator have been multiplied by the
same expressions.
Introduce now a bosonic Vandermonde determinant:
\begin{eqnarray}
\label{B2} {\cal V}_s(\{x_1, \dots x_s\})= \prod_{i<j=1}^s
(x_i-x_j)=Det \left[
\begin{array}{cccc}
1 & x_1 & \dots & x_1^{s-1}\\
1 & x_2 & \dots & x_2^{s-1}\\
\vdots & \vdots & \vdots & \vdots\\
1 & x_s & \dots & x_s^{s-1}\\
\end{array}
\right]
\end{eqnarray}
The expression (\ref{B1}) for the partition function can be rewritten as
\begin{eqnarray}
\label{B3} Z_{n,N}(\epsilon)=\lim_{\{e_{\alpha}\}\rightarrow
e}\sum_{\{{\sigma}\}} (-1)^{P} \frac{{\cal
V}_{n+N}(\{\theta_1,\dots \theta_{n+N}\}) {\cal
V}_{2M-N}(\{\mu_{\bar{\sigma}(1)},\dots
\mu_{\bar{\sigma}(2M-N)}\})}{{\cal V}_{n}(\{e_1, \dots e_n
\}){\cal V}_{2M}(\{\mu_1,\dots \mu_{2M}\})},
\end{eqnarray}
where $\{\theta_1, \dots \theta_{n+N}\} = \{e_1, \dots e_{n},
\mu_{\sigma(1)},\dots \mu_{\sigma(N)}\}$ and $(-1)^P$ is the
parity of permutation
\begin{equation}
\label{P} P:\{e_1, \dots e_n, \mu_1, \dots \mu_{2M}\}\rightarrow
\{e_1, \dots e_n, \mu_{\sigma(1)},\dots \mu_{\sigma(N)},
\mu_{\bar{\sigma}(1)},\dots \mu_{\bar{\sigma}(2M-N)}\}.
\end{equation}
 The denominator in the expression (\ref{B3}) does not depend on
$\sigma$ and can be taken out of summation, while $\sum$ of
nominators is nothing but the determinant of the matrix $A_0$:
\begin{eqnarray}
\label{B4}
A_0=\left(%
\begin{array}{ccccccccc}
  1 & e_1 & e_1^2 & \dots & e_1^{n+N-1} & 0 & 0 & \dots & 0 \\
  1 & e_2 & e_2^2 & \dots & e_2^{n+N-1} & 0 & 0 & \dots & 0 \\
  \vdots & \vdots & \vdots & \vdots & \vdots & \vdots & \vdots & \vdots & \vdots \\
  1 & e_n & e_n^{2} & \dots & e_n^{n+N-1} & 0 & 0 & \dots & 0 \\
  1 & \mu_1 & \mu_1^{2} & \dots & \mu_1^{n+N-1} & 1 & \mu_1 & \dots & \mu_1^{2M-N-1} \\
  \vdots & \vdots & \vdots & \vdots & \vdots & \vdots & \vdots & \vdots & \vdots \\
  1 & \mu_M & \mu_M^{2} & \dots & \mu_M^{n+N-1} & 1 & \mu_M & \dots & \mu_M^{2M-N-1} \\
  0 & 0 & 0 & \dots & 0 & 1 &\mu_{M+1} & \dots & \mu_{M+1}^{2M-N-1} \\
  \vdots & \vdots & \vdots & \vdots & \vdots & \vdots & \vdots & \vdots & \vdots \\
  0 & 0 & 0 & \dots & 0 & 1 & \mu_{2M} & \dots & \mu_{2M}^{2M-N-1} \\
\end{array}%
\right),
\end{eqnarray}
calculated by $N+n$ and $2M-N$ dimensional minors and by use of
the formula (\ref{B2}).

Now let us consider the operator $\delta_x = x {\partial \over
\partial x}$ introduced earlier and observe that
\begin{equation}\label{delta}
e^{-x}(\delta_x)^p e^x = x^p  + {\cal P}^{(p-1)}(x)
\end{equation}
where ${\cal P}^{(p-1)}(x)$ is the polynomial of $p-1$ order of
$x$. From this it follows, that if we replace the monomials
$x^p$ in the matrix $A_0$ with $e^{-x}(\delta_x)^p e^x $, we will
obtain the matrix $A$ defined in (\ref{det} ). Their
determinants relate as
\begin{eqnarray}
det[A_0] = det[I^{-1} A]
\end{eqnarray}
and define the nominators in the expressions (\ref{Z12}) and
(\ref{B3}) for the partition function $Z_{n,N}(\epsilon)$. The
notation $I=\exp(\sum_{\alpha=1}^n e_s +\sum_{k=1}^{2M}\mu_k)$
is defined just after formula (\ref{Z12}).

The next step is the consideration of the degenerate limit
$e_{\alpha} \rightarrow e; \; \alpha =1, \dots ,n$.
We put $e_1=e$ and consider Taylor expansions
$f(e_{\alpha})= f(e) + (e_{\alpha}-e)\partial_e f(e)+ \dots
{(e_{\alpha}-e)^{n-1} \over (n-1)!}\partial_e^{n-1} f(e)+\dots$
around $e$ of all entities in the
upper-left $n \times n$ corner of the matrix $A$ in (\ref{Z12}).
It is easy to see that in the limit $e_{\alpha} \rightarrow e; \; \alpha =1,
\dots ,n$,
higher than $n-1$ orders of the expansions will not contribute into the matrix
$A$, while the coefficients $(e_{\alpha}-e)^k$ will form a Vandermonde determinant
(\ref{B2}), which will cancel the same ${\cal V}_{n}(\{e_1, \dots ,e_n
\})$ in the denominator of $Z_{n,N}(\epsilon)$ in the formula (\ref{B3}).

Therefore we obtain
\begin{eqnarray}
\label{B5}
Z_{n,N}(\epsilon) = \frac{(2 \pi)^N (C_M)^N i^{nN}}{{\cal V}_{2M}
(\{\mu_1,\dots \mu_{2M}\})} I^{-1} det[B_{n,N}],
\end{eqnarray}
with $I=\exp(n e + \sum_r \mu_r)$ and $B_{n,N}$ equal to
\begin{eqnarray}
\label{B6}
\left (
\begin{array}{ccccccc}
I_1 & \delta_1I_1 &  \dots & \delta_1^{n+N-1} I_1 & 0 & \dots & 0\\
\frac{1}{1!}\partial_eI_1 & \frac{1}{1!}\partial_e\delta_1I_1 &  \dots & \frac{1}{1!}
\partial_e\delta_1^{n+N-1} I_1 & 0 & \dots & 0\\
\vdots & \vdots &  \vdots & \vdots &  \vdots & \vdots & \vdots \\
\frac{1}{(n-1)!}\partial_e^{n-1}I_1 & \frac{1}{(n-1)!}\partial_e^{n-1}\delta_1I_1 & \dots &
\frac{1}{(n-1)!}\partial_e^{n-1}\delta_1^{n+N-1} I_1 & 0 &  \dots & 0\\
\bar{I}_1 & \bar \delta_1 \bar{I}_1 & \dots & \bar
\delta_1^{n+N-1} \bar{I}_1 &  \bar{I}_1  &
\dots &\bar \delta_1^{2M-N-1} \bar{I}_1\\
\vdots & \vdots & \vdots & \vdots &  \vdots & \vdots & \vdots \\
\bar{I}_M & \bar \delta_{M} \bar{I}_M &  \dots & \bar
\delta_{M}^{n+N-1} \bar{I}_M &  \bar{I}_M
& \dots &\bar{\delta}_{M}^{2M-N-1} \bar{I}_M \\
0 & 0 &  \dots & 0 &  \bar{I}_{M+1}
& \dots &  \bar{\delta} _{M+1}^{2M-N-1} \bar{I}_{M+1} \\
\vdots & \vdots & \vdots  &  \vdots & \vdots &  \vdots & \vdots \\
0 & 0 &  \dots & 0 &  \bar{I}_{2M} & \dots
& \bar{\delta}_{2M}^{2M-N-1} \bar{I}_M \\
\end{array}
\right).
\end{eqnarray}
In this expression $I_1 = \exp(e)$ and $\delta_1 = e \partial_e$.

In order to derive the Toda lattice equation (\ref{GTLE}) we will
follow the technique presented in the article \cite{SV2} in
connection with QCD partition function. The method is based on use
of the Sylvester identity \cite{Syl}. For the determinant of any
given matrix $D$
\begin{equation}
\label{Syl}
C_{i,j}C_{p,q}-C_{i,q}C_{p,j}= det[D]C_{i,j;p,q},
\end{equation}
where $C_{i,j}$ is the cofactor of the matrix element $ij$:
\begin{equation}
\label{Syl2}
C_{i,j} = \frac{\partial det[D]}{\partial D_{i,j}},
\end{equation}
and $C_{i,j;p,q}$ is the double cofactor of the matrix elements $ij$
and $pq$:
\begin{equation}
\label{Syl3}
C_{i,j;p,q} = \frac{\partial^2 det[D]}
{\partial D_{i,j}\partial D_{p,q}}.
\end{equation}

Let us consider now the $(2M+n+1)\times(2M+n+1)$ dimensional matrices
$B_{n+1,N}$ and $B_{n+1,N-1}$
and write Sylvester identities (\ref{Syl}) for different set of
indices ${i,j;p,q}$ for each of them.

For the $B_{n+1,N}$ we  take $i= n+1,\; j=n+N+1,\;
p= n,\; q= n+N$  and obtain
\begin{eqnarray}
\label{Syl4}
C_{n+1,n+N+1}C_{n,n+N}&-&C_{n,n+N+1}C_{n+1,n+N}\nonumber\\
&=& det[B_{n+1,N}]C_{n+1,n+N+1;n,n+N}.
\end{eqnarray}

From analogous to (\ref{B6}) structure of $B_{n+1,N}$
%From the structure (\ref{B6}), corresponding to $B_{n+1,N}$, it
follows that
\begin{eqnarray}
\label{B7}
C_{n+1,n+N+1}= det[B_{n,N}], \;\qquad \quad
C_{n,n+N+1}= \frac{1}{n}\partial_e det[B_{n,N}],\nonumber\\
C_{n+1,n+N+1;n,n+N}= det[B_{n-1,N}], \qquad \qquad \qquad \\
C_{n,n+N} = \frac{1}{n}\partial_e det[B_{n,N}^{\prime}], \;\qquad \quad \quad
C_{n+1,n+N}= det[B_{n,N}^{\prime}],\qquad \nonumber
\end{eqnarray}
where the matrix $B_{n,N}^{\prime}$ is defined by acting
by the operator
$\delta^{1+2M}=e \frac{\partial}{\partial e} + \sum_{r=1}^{2M}\mu_r
\frac{\partial}{\partial \mu_r}$
on the $(n+N)$-th column of the matrix $B_{n,N}$.
% by the operator
%$\delta^{1+2M}=e \frac{\partial}{\partial e} + \sum_{r=1}^{2M}\mu_r
%\frac{\partial}{\partial \mu_r}$.

Now consider the Sylvester identity for the matrix $B_{n+1,N-1}$.
Setting  $i= n+1,\; j=n+2M+1,\; p= n,\; q= n+2M, $ we will have
%Second setting  $i= n+1,\; j=n+2M+1,\;
%p= n,\; q= n+2M $ we take for the matrix $B_{n+1,N-1}$
%and write Sylvester identity
\begin{eqnarray}
\label{Syl5}
C_{n+1,n+2M+1}^{\prime}C_{n,n+2M}^{\prime}&-&C_{n,n+2M+1}^{\prime}
C_{n+1,n+2M}^{\prime}\nonumber\\
&=& det[B_{n+1,N-1}]C_{n+1,n+2M+1;n,n+2M}^{\prime}.
\end{eqnarray}

As in the previous case and from the structure of (\ref{B6})
for $B_{n+1,N-1}$ it
follows that
\begin{eqnarray}
\label{B71}
C_{n+1,n+2M+1}^{\prime}= det[B_{n,N}], \;\qquad \quad
C_{n,n+2M+1}^{\prime}= \frac{1}{n}\partial_e det[B_{n,N}],\nonumber\\
C_{n+1,n+2M+1;n,n+N}^{\prime}= det[B_{n-1,N+1}], \qquad \qquad \qquad \\
C_{n,n+N}^{\prime} = \frac{1}{n}\partial_e det[B_{n,N}^{\prime\prime}],
\;\qquad \quad \quad
C_{n+1,n+N}^{\prime}= det[B_{n,N}^{\prime\prime}],\qquad \nonumber
\end{eqnarray}
where the matrix $B_{n,N}^{\prime\prime}$ is defined by acting
by the operator $\delta^{1+2M}$ on the
$(n+2M)$-th column of the matrix $B_{n,N}$.

It is easy to observe, that
\begin{equation}
\label{B9}
det[B_{n,N}^{\prime}]+det[B_{n,N}^{\prime\prime}]= \delta^{1+2M}det[B_{n,N}].
\end{equation}

Inserting now expressons (\ref{B7}) and (\ref{B71}) into the Sylvester
identities (\ref{Syl4}) and (\ref{Syl5}) respectively, taking their
sum and using (\ref{B9}) we get
\begin{eqnarray}
\label{B8}
det[B_{n,N}]^2 \frac{\partial}{\partial e}\delta^{1+2M}\ln[det[B_{n,N}]]&=&\\
= n\; det[B_{n+1,N}]det[B_{n-1,N}]&+&n\; det[B_{n+1,N-1}]det[B_{n-1,N+1}].\nonumber
\end{eqnarray}
Finally, using the relation (\ref{B5}) between the partition function
and the matrix $B_{n,N}$ we obtain the desired graded Toda lattice
equation as
\begin{eqnarray}
\label{GTLE2} n+\partial _{e}\delta ^{1+M}\ln
[Z_{n,N}(\epsilon)]&=& \nonumber \\
n\frac{Z_{n+1,N}(\epsilon)Z_{n-1,N}
(\epsilon)}{Z_{n,N}^2(\epsilon)}&+&n\frac{Z_{n+1,N-1}(\epsilon)Z_{n-1,N+1}
(\epsilon)}{Z_{n,N}^2(\epsilon)},
\end{eqnarray}
which coincides with (\ref{GTLE}) after substitution $e=i
\epsilon$. Notice that the Vandermonde determinants, ${\cal
V}_{2M}(\{\mu_1,\dots \mu_{2M}\})$, do not contribute to the left
hand side of the Toda lattice equation since
$\delta^{1+2M}\ln[{\cal V}_{2M}(\{\mu_1,\dots \mu_{2M}\})]$ is a
number, while the contribution of  $I=\exp(n e + \sum_{r=1}^{2M}
\mu_r)$ gives addition of $n$.

If we take $N=M$, then, since $B_{n-1,N+1}=0$, the graded Toda equation
(\ref{GTLE2}) simplifies.

\end{document}